# Enhancing triboelectric nanogenerators power conversion efficiency with few-layers graphene flexible electrodes

G. Pace,[1,2*] A. Ansaldo,[1] M. Serri,[1] S. Lauciello,[1] and F. Bonaccorso[1,3,*]

[1]*Istituto Italiano di Tecnologia, Graphene Labs, Via Morego, 30, 16136 Genova, Italy*
*Institute for Microelectronics and Microsystems, National Research Council, Via C. Olivetti 2, 20864 Agrate (Milan), Italy*
[3] *BeDimensional S.p.A., via Albisola 121, 16163 Genova, Italy*

Corresponding authors: *giuseppina.pace@iit.it*,
*francesco.bonaccorso@iit.it*

**Abstract**

The development of portable and wearable electronics requiring only few tens of microwatts to operate fosters the search for low power energy sources. Moreover, the fast-growing demand for new sustainable energy has raised the interest in energy harvesters able to convert mechanical energy into electrical power. Triboelectric nanogenerators (TENGs) can provide such green power supply. Recent developments in TENGs technology show that they can be used in different applications ranging from wearable self-powered sensors to wind and sea wave energy harvesting. In spite of the wide number of TENGs developed so far, an in depth understanding of their working principle is still missing. In this work, we highlight the fundamental role played by the interface between the triboelectric material and the electrode collector in contributing to the TENG's power generation. We show that by simply replacing the "standard" gold electrodes with flexible few layer graphene electrodes in TENGs operating in vertical contact mode, a 26-fold increase in power density is achieved. Here, we elucidate the main mechanism at the base of such boost in power output, describing guidelines for the combination of electrode and triboelectric materials to optimize the mechanical energy conversion efficiency in TENGs.

## 1. Introduction

The field of flexible and wearable electronics[1] is boosting the development of low cost, lightweight devices and sensors based on solution processable conductors[2,3] and semiconductors.[4,5] However, to achieve the full integration of flexible electronics into wearables it is necessary to develop portable power sources required for their operation. For this purpose, self-powered electronics able to convert the green environmental energy sources into a readily-available power supply are needed. For triggering and operating a wide number of smart electronic devices, as the ones optimized for the Internet of Things (IoT) and wearable electronics, only a low power output is required (tens of $\mu W/cm^2$).[9] A recent research field, which focuses on the development of triboelectric nanogenerators (TENGs),[6] also defined as kinetic energy harvester,[7] aims at converting the mechanical energy present in nature, and otherwise wasted in the environment, into electrical power.[8] Triboelectric nanogenerators have already shown to be able to convert mechanical energy from green and renewable sources such as human body motion,[10,11,12,13] wind[14,15] and sea waves,[16] providing an electrical power output ranging from few tens of $\mu W/cm^2$ up to few tens of $mW/cm^2$.[17,18,19] Their development as green power harvesters would make electronic devices independent from external power recharging, enabling the development of self-powered electronics and sensors. The variability in TENGs power density is due to the use of different materials and devices structure, as well as to the different environmental mechanical energy that can be harvested with TENG. The latter is characterized by different load forces operating at diverse frequencies.[20] In particular, the mechanical-to-electrical energy conversion efficiency increases with increasing the applied load and mechanical input frequency.[21]

Although triboelectric materials have been investigated for long time in the field of electrets[22,23,24] and tribology,[25] a renewed attention for such materials has raised in the last decade due to the development of TENGs.[26]

Such nanogenerators exploit the charge transfer occurring between two triboelectric materials, *i.e.*, between materials that have strong tendency to charge electrostatically upon physical contact. The selection of the insulating tribomaterials that are integrated in a TENG, is generally based on the triboelectric series,[27, 28] in which the materials are ordered according to their different tendency to gain (electronegative materials) or lose (electropositive materials) electrons.[28,29] Ideal candidates to be integrated into TENGs are therefore those materials placed at the opposite side of the triboelectric series, which own very different electron affinity.[25] Triboelectric nanogenerators can be classified in four basic modes: vertical contact-separation mode,[30] in plane contact-sliding mode,[31] single-electrode mode,[32] and freestanding triboelectric-layer mode.[33] The most common configuration is the vertical contact-separation mode, in which each of the two tribomaterials is interfaced with an electrode acting as charge collector.[18] The combined tribomaterial and interfacing electrode will be hereon defined as triboelectrode. An external circuit ensures the electrical connection between the two triboelectrodes (Figure 1a).

In order to enable TENGs cost-effective production by means of printable technologies, such as roll-to-roll coating, TENGs' active materials have to be solution processable.[35,36] A major constrain to achieve a full solution processability, and therefore to lowering the production costs of such nanogenerators, arises from the deposition techniques used to fabricate the electrodes, which often rely on ultra-high vacuum and high temperature processing.[37] While a rationale can be followed for the proper selection and processing of the tribomaterials, there are no insights on how to perform the best electrode choice for TENGs fabrication. So far, different electrodes have been used in TENGs including metals,[30] carbon nanotubes,[38] graphene,[39] and PEDOT-PSS[40] based electrodes. Nevertheless, the main focus was mostly centered on finding the best electrode for achieving different properties such as flexibility[41] and excellent mechanical[42] and chemical[43] stability. In this context, previous works have already shown the integration of transparent and flexible graphene-based electrodes in TENGs.[44,45] However, those electrodes were obtained via transfer printing on flexible substrates of a graphene film grown on Cu substrate via chemical vapor deposition (CVD),[39,46–48] which is not suitable for large area and low-cost applications.[49] In this work, we elucidate the principles that make few-layer graphene-based flexible electrodes a better choice over evaporated metal electrodes. The few-layer graphene flakes prepared by wet-jet-milling[50] were used as starting material for the formulation of few-layer graphene-based pastes deposited on flexible substrates via doctor blading in ambient condition (Figure 1b). Few-layers graphene adds to its solution processability a high charge carriers mobility,[51] which makes it a versatile material for the fabrication of conducting electrodes[52,53] to be applied in different technologies, ranging from energy harvesting[54] and generation[55] to field effect transistors[56] and sensors.[57–59]

Thought triboelectric charging is almost ubiquitous in our daily life, as excess electrostatic charges are present on almost all dielectric materials including fabrics and plastics, there is still an open debate,[25,60] both on the mechanism associated to the charge transfer between two triboelectric materials (tribomaterials) and on the parameters that control the charge accumulation and the induction current in TENGs. In the current work, we contribute to elucidate the criteria to be followed for the appropriate selection of the electrodes in TENGs to the purpose of optimizing the mechanical-to-electrical power conversion efficiency. TENGs based on few-layers graphene (FLG) flexible electrodes have been compared with Au electrodes evaporated onto flexible polyethylene naphthalate (PEN) substrates. When using the FLG electrodes in TENGs a 26-fold increase in power density (28.6 $\mu W/cm^2$ at 100MΩ) is reached with respect to TENG fabricated with gold electrodes (1.1 $\mu W/cm^2$ at 100MΩ). Based on this experimental evidence, we demonstrate the fundamental role played by the interface between the triboelectric material and the electrode collector. We describe how interface parameters such as the electrodes surface morphology, work function and interfacial dipoles can be tuned to achieve the maximum mechanical power conversion efficiency in TENGs. The role played by the electrode work function is assessed by fabricating TENGs incorporating Au electrodes functionalized with different self-assembled monolayers (SAMs). Based on the observed TENGs performance dependence on the electrode work function, we provide a rationale to optimize the combination of electrode work function and tribomaterials.

2. **Materials and Methods**

Gold electrodes were prepared by e-beam deposition of 5 nm Ti and 70 nm Au on polyethylenetherepthalate (PEN 50 µm thick, Dupont-Tejin) substrates.

Few-layer graphene flakes were produced via wet-jet-milling (WJM) as previously described.[34] The defect-free and high quality 2D-crystal dispersions was then dried under vacuum and made into powder.[50]

The flexible FLG electrodes were prepared from a slurry composed of few-layer graphene flakes (WJM), carbon black (Super P, Alfa Aesar), ethylene vynil acetate (Eva, Mr Watt Srl) with a weight ratio of 1:250:40 dissolved in a solvent mixture containing butylcarbamate and chlorobenzene (3:7.5 v/v). The slurry was coated on a clean Al foil, through the doctor blade technique and dried under ambient condition at 80°C overnight. The few-layer graphene-based film was transferred onto Eva flexible substrates by pressing the few-layer graphene coated Al film on Eva flexible substrates at a temperature of 80°C for 15 min. The Al foil was than peeled off the FLG electrode prior to use. The FLG electrode uniformly covers the entire area, showing an average thickness of 5 µm.

Nylon membranes (0.2 µm pores diameter; 200 µm thickness) and PVDF membranes (0.2 µm pores diameter; 200 µm thickness) were supplied by Merk and Ge Healthcare (Amersham), respectively. Polyurethane membranes (40 µm thickness) were purchased from Sigma. The device active area was fixed at a value of $4 \times 4$ cm$^2$.

Field-emission scanning electron microscope FE-SEM (JEOL JSM-7500 FA) was used to acquire the SEM data with the acceleration voltage set to 5 kV. The flexible few-layer graphene/polyurethane triboelectrodes were frozen and then broken in a liquid N$_2$ tank before the SEM cross section images were acquired.

The X-ray photoelectron spectroscopy (XPS) analysis is accomplished on a Kratos Axis UltraDLD spectrometer at a vacuum better than 10$^{-8}$ mbar, using a monochromatic Al Kα source operating at 20 mA and 15 kV and collecting photoelectrons from a $300 \times 700$ µm$^2$ sample area. The charge compensation device was not used. Wide spectra were acquired at pass energy of 160 eV and energy step of 1 eV, while high-resolution spectra of O 1s, N1s, C 1s and Au 4f peaks were acquired at pass energy of 10 eV and energy step of 0.1 eV. The samples were mounted on the sample holder with copper clips and quickly transferred from air to the XPS chamber. Data analysis is carried out with CasaXPS software (version 2.3.19PR1.0). The energy scale was calibrated by setting the Au 4f$^{7/2}$ peak at 84.0 eV and C 1s peak (graphitic component) at 284.3 eV.

Ultraviolet photoelectron spectroscopy (UPS) with He I (hν = 21.2 eV) radiation was performed to estimate the Fermi energy level (E$_F$) and the valence band maximum of the materials under investigation. The experiments were conducted on the samples after the XPS analysis using the same equipment. Spectra were acquired at pass energy of 10 eV and energy step of 25 meV, collecting photoelectrons from a spot with 55 µm diameter. A −9.0 V bias was applied to the sample in order to precisely determine the low kinetic energy cut-off. The energy scale was corrected according to the binding energy calibration performed for the XPS measurement.

Conductivity of the FLG flexible electrode was measured with the van der Pauw method on a 4×4 cm² electrode. We also measured the sheet resistance of the pastes bar coated on kapton substrates and of gold electrodes on PEN with a four-point probe Jandel system (RM300). We measured a sheet resistance of (142 ± 5) Ω/□ for the FLG flexible electrode and (1.1 ± 0.2) Ω/□ for the thin film gold. The sheet resistance of the FLG electrodes does not vary with the addition of the carbon black (139 ± 5 Ω/□).

Self-assembled monolayers on gold were prepared by overnight immersion of the clean gold substrates in 1 mM solution in ethanol. After incubation, the substrates were rinsed with ethanol and dried under $N_2$. 1-dodecanethiol and aminoethanthiol hydrochloride were purchased from Sigma.

The triboelectric characterization of the TENGs was performed using a shaker setup (TIRA BAA 60), an oscilloscope (Tecktronix MSO5000), a voltage probe 40 MΩ (Tecktronics) for voltage measurements and a current amplifier (1211 DL Instruments) for current measurements. A home built resistance commutator was used during power measurements. The power density was calculated from the root mean square (RMS) value of the current measured during the press and release cycle. The time dependent sinusoidal force profile is reported in Figure SI-8.

### 3. Results and discussion

The energy generation in TENGs is based on the Maxwell electrostatic induction and displacement current,[34] which follow the charge transfer and surface electrification occurring upon contact of two materials with different electrons affinity and ionization potential (triboelectrification).[61,62,63,64,60,65,66] The Maxwell displacement current ($J_D$), which is dependent on the displacement (D), electric (E) and polarization (P) field, with ε being the permittivity, is given by the following equation:[34]

$$J_D = \frac{\partial D}{\partial t} = \varepsilon \frac{\partial E}{\partial t} + \frac{\partial P}{\partial t} \qquad (1)$$

In this work, the experimental configuration used for testing TENGs is the vertical contact-separation mode,[18] in which two triboelectrodes are placed in parallel one with respect to the other, being initially kept separated by an air gap (Figure 1a).[21] Under operation conditions the distance between the two tribomaterials in the TENG is varied in a cyclic way, *i.e.*, the two tribomaterials are brought into contact and subsequently separated with a sequence of press and release actions.[21] Upon contact of the two triboelectric materials, due to their different charge affinity, an interfacial charge transfer occurs and trapped excess charges accumulate at each tribomaterial surface.[60, 66] In the following separation step, the electrostatic field originated by the triboelectric charges triggers a charge flow through the external circuit and load [34] (see supporting information, SI-1). The driving force for the generation of the induction current is therefore the nullification of the electric potential that is generated between the electrodes due to the presence of triboelectric charges accumulated at the tribomaterial surfaces upon separation.[34] The amount of charge needed to maintain the electrodes at the same potential varies with the relative distance between the two triboelectrodes, enabling the current to flow between the two triboelectrodes across the external circuit during the press and release operation.[34]

Different electrodes and triboelectric materials have been explored for TENGs fabrication, but only few examples can be found in which solution processable conductors are embedded into TENGs.[67,68] In this work, we fabricated FLG flexible electrodes, using all wet and ambient condition processing. We used few-layer graphene flakes prepared by wet-jet-milling[50] to obtain a graphene paste that was deposited as a film onto an ethylene-vinyl-acetate (EVA) flexible substrate by doctor blading (experimental and scheme Figure 1b), a technique that can be easily scaled up for larger production processing.

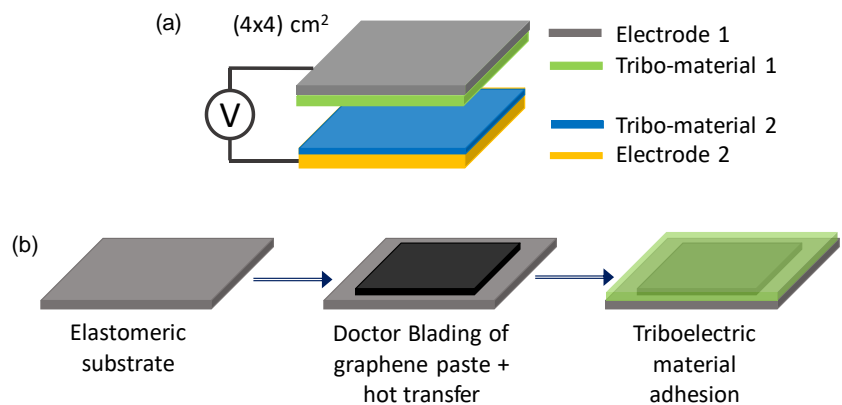

**Figure 1:** (a) Structure of a triboelectric nanogenerator operating in vertical contact composed of top and bottom triboelectrodes. (b) Fabrication steps for the graphene-based flexible triboelectrode. Tribomaterial 1 and tribomaterial 2 refer to the two triboelectric materials owing different charge affinity.

For this study we selected polyammide 6,6 (nylon) and polyvinylfluoride (PVDF), which are known to be at the opposite side of the triboelectric series, being an electropositive and electronegative tribomaterial, respectively.[29] Since the interfacial charge transfer occurs upon contact between the two tribomaterials, increasing their surface area ensures a higher probability for interfacial charge transfer and surface electrification.[69] To increase the tribomaterial surface area we used porous nylon (200 µm thick) and PVDF membranes (200 µm thick, SEM images available in SI-2). Each tribomaterial has been interfaced with either e-beam evaporated gold electrodes or FLG flexible electrodes to form a triboelectrode. Figure 2-a show the cross-section SEM images of the inner structure of the FLG flexible triboelectrode, in which the top layer is composed of the EVA flexible substrate, while the bottom layer is composed of a polyurethane film. Polyurethane was used for acquiring SEM cross section images due to its stability under freeze cut in liquid nitrogen.[34] Figure 2-b shows the mesoscopic structure of the conductive graphene interlayer, whose average thickness is 5 µm and is composed by graphene flakes, carbon black nanoparticles and polymer binder (EVA). The addition of the carbon black nanoparticles does not influence the electrode conductivity (see Experimental section).

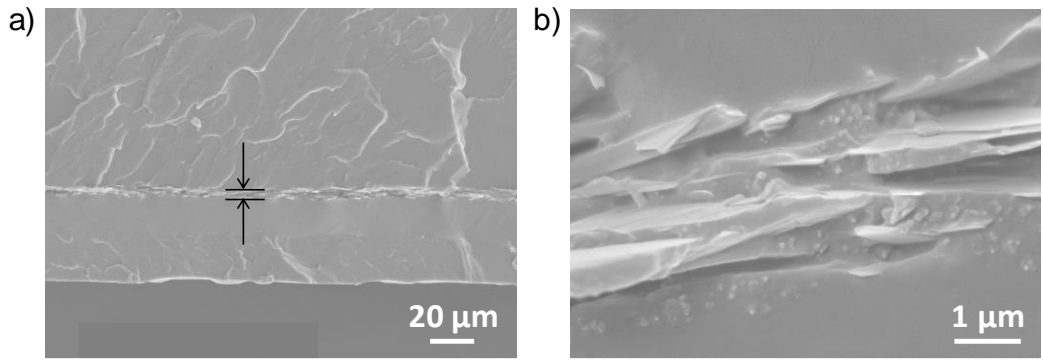

**Figure 2:** SEM cross section images of graphene-based flexible triboelectrode. a) Large scale image showing the top layer of EVA plastic flexible substrate, the bottom polyurethane triboelectric material and the intermediate FLG electrode. b) Zoom on the FLG electrode layer where graphene flakes and carbon nanoparticles can be observed.

It is well known,[21] that the power generated by TENGs depends not only on the properties of the two tribomaterials, as described above, but also on the device operational parameters, namely the force applied during the press and release cycle, frequency and dynamic of the mechanical motion.[70] To be integrated in wearable electronics, TENGs should operate under the low frequency movement generated by the human body. In this work, we tested the developed TENGs under force and frequency parameters that find application in wearable technologies. In particular, we operated all the devices under the same frequency (3 Hz), force (10 N) and maximum separation distance (5 mm).[71]

In Figure 3 we compare TENGs containing nylon and PVDF membranes and either gold electrodes (Au/PVDF // Nylon/Au) or FLG electrodes (graphene/PVDF // Nylon/graphene). The electrical conductivity of the electrodes, measured with the van der Pauw method,[72] reported a value of $4.4 \times 10^3$ S/m for the FLG electrodes and $2.2 \times 10^7$ S/m for the gold ones, in agreement with literature.[73,74,75–77,78] The electrical stability of the FLG flexible electrodes under continuous bending was also tested and reported in the SI (Figure SI-3). In spite of the almost 4 order of magnitude difference in electrical conductivity, the power density generated by TENGs incorporating the FLG electrodes (28.6 µW/cm$^2$, Figure 3e) show a 26-fold increase in power density compared to TENGs based on the Au electrodes (1.1 µW/cm$^2$, Figure 3b). Since we use the same tribomaterials in both TENGs, *i.e.*, the PVDF and nylon membranes, these data demonstrate that a higher charge density is accumulated at the interface between the FLG electrode and tribomaterial (~ 2nC/cm$^2$) compared to gold electrodes (~ 0.38 nC/cm$^2$), despite the significant difference in the measured electrical conductivities.

This experimental evidence is a first demonstration of the crucial role played by the interface between the active material and the electrode in contributing to the power density in TENGs.

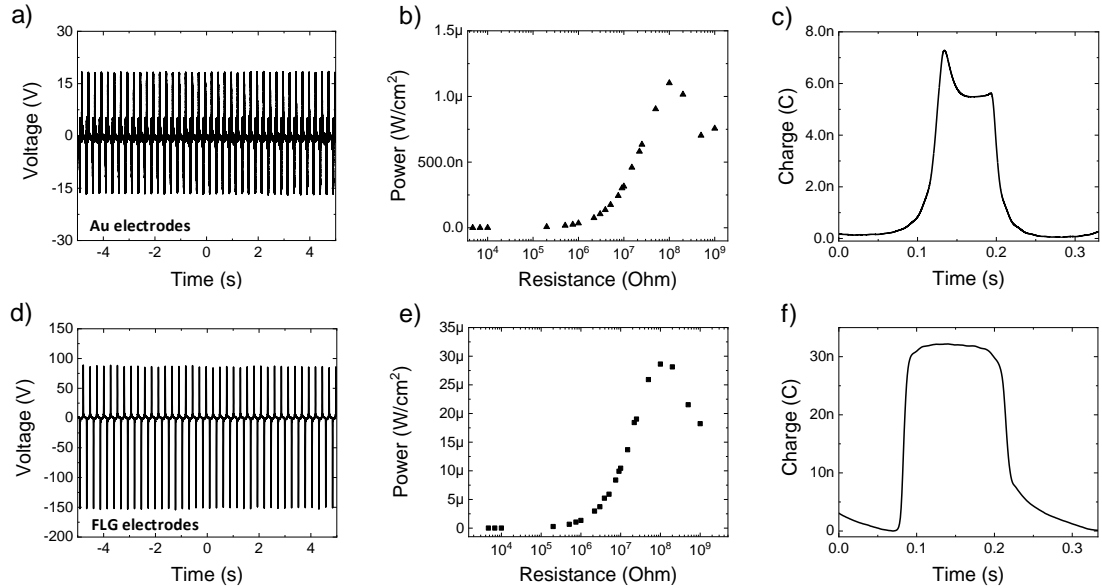

**Figure 3:** a) Open circuit voltage, b) power efficiency and c) transferred charges measured on a TENG including a gold electrode (Au/PDVF // Nylon/Au). d) Open circuit voltage and e) power efficiency and f) transferred charges measured on a TENG consisting of only FLG flexible electrodes (graphene/PDVF // Nylon/graphene). (40 MΩ probe, 10 N force, 5 mm, 4 Hz).

To better understand the mechanism behind the higher charge density accumulated at the FLG electrode/tribomaterial interface compared to the gold electrodes, we acquired SEM images of the two electrode surfaces (Figure SI-4).

Due to the presence of carbon nanoparticles agglomerates and irregularly distributed few-layer graphene flakes, the FLG electrode shows a rougher surface compared to the gold ones. Therefore, a higher electrode capacitance can be one reason at the base of the observed increase in power density generated in TENGs based on FLG electrodes, compared to the gold-based ones.[79] We further investigated the role played by the electrode/tribomaterial interface, comparing TENGs incorporating either bare Au electrodes or Au electrodes functionalized with different thiolated self-assembled monolayers (1-dodecanethiol SAM, Au-$CH_3$; 1-aminoethanethiol SAM, Au-$NH_2$, Figure SI-5). The functionalization of gold surfaces with SAMs is well known to vary the Au work functions through the addition of chemisorbed molecular dipoles.[80,81] We extracted the work function values of each electrode from UPS measurements (Figure 4), which are 4.88 eV, 4.66 eV, 4.37 eV for bare Au, Au-$NH_2$, Au-$CH_3$ respectively, and 4.20 eV for flexible FLG electrodes.

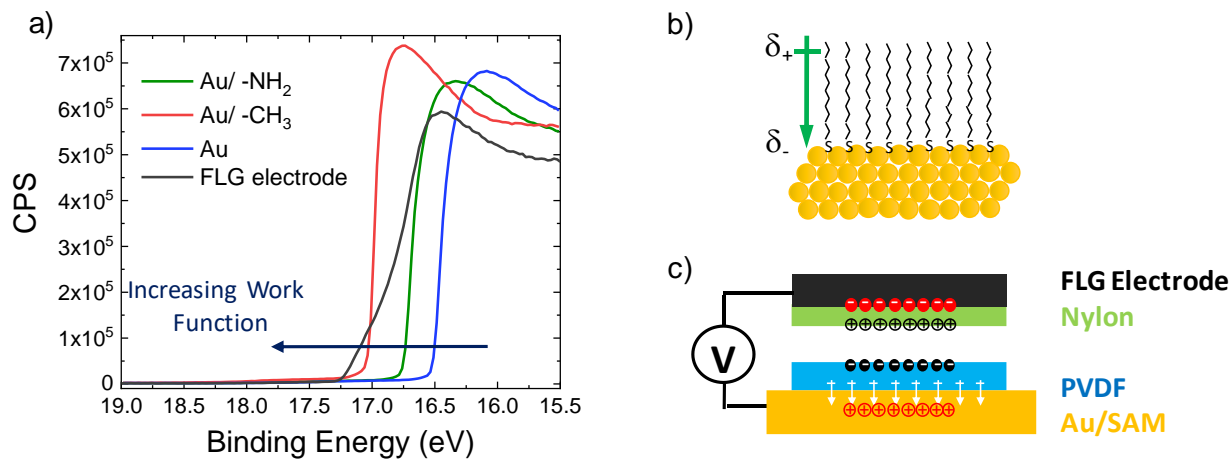

**Figure 4:** a) UPS data acquired on the flexible FLG electrodes and Au electrodes with different surface functionalization. b) Dipole moment orientation introduced on top of the Au surface upon alkylthiol SAMs formation. c) Schematic of the favorable orientation of the alkylthiol SAM's dipole placed at the interface between the Au electrode and a negative triboelectric material such as PVDF.

A shift towards lower work function values can be observed when comparing the bare Au electrodes with the SAMs modified Au ones,[80] and an even lower value is found for the flexible FLG electrodes. The decreasing work function observed with the SAMs modified electrode, is compatible with the expected surface dipole orientation associated to the addition of alkyl and aminoalkyl thiolated molecules (Figure 4b).[81]

Figure 5 shows that the TENGs power density increases when the thiolated SAMs are chemisorbed on the Au surface and interfaced with the negative tribomaterial (PVDF) (3.8 $\mu W/cm^2$ for Au-CH$_3$; 1.6 $\mu W/cm^2$ for Au-NH$_2$; 1.1 $\mu W/cm^2$ for Au). A ~3.5-fold improvement of the TENG power density is obtained with the Au-CH$_3$ functionalized electrode, while lower improvement (~1.5-fold increase) is obtained with the Au-NH$_2$ electrode compared to the bare Au electrode. Though surface contact area is an important parameter to be considered in charging capacitors, the results obtained with functionalized Au electrodes demonstrate that the contact area is not a unique player in determining the increasing charge density accumulation at the electrode/tribomaterial interface.

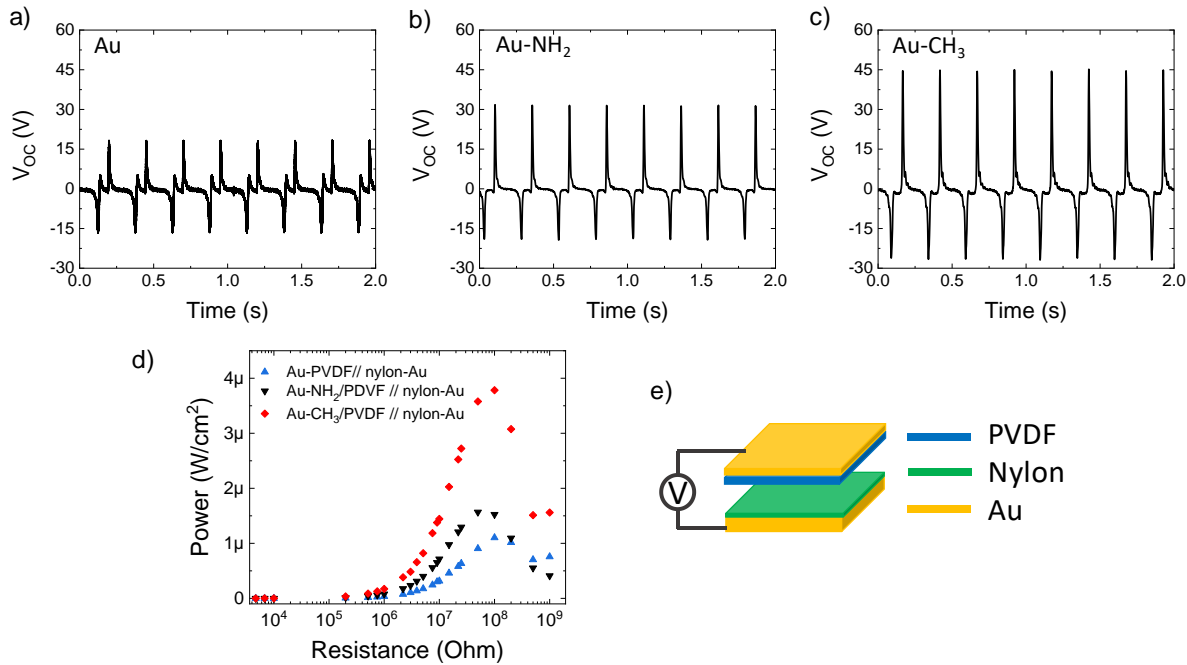

**Figure 5:** a-c) Open circuit voltage and d) power density of TENGs based on gold electrodes with different surface functionalization: a) bare Au; b) 1-dodecanthiol SAM on gold (Au-CH$_3$); c) 1-aminoethanthiol SAM on gold (Au-NH$_2$); e) schematic of Au-TENGs. (40MΩ, 10N, 5mm, 4Hz).

We also verified the impact of the SAM functionalized Au electrodes on TENGs, when those electrodes are interfacing a positive triboelectric material (Nylon, Figure SI-6). For these devices we recorded a small decrease in the devices power density (1.0 µW/cm$^2$ for Au-CH$_3$; 1.0 µW/cm$^2$ for Au-NH$_2$; 1.1 µW/cm$^2$ for Au, Figure SI-6) compared to the values reported in Figure 3, in which the electrodes were interfacing a negative tribo-material (PVDF). This experimental evidence shows that the influence of surface engineering over TENGs performance is also dependent on the polarization of the insulating tribomaterial upon contact electrification. Thus, the selection of the electrode collectors is strictly relying also upon the choice made on positive (*e.g.*, Nylon) or negative (*e.g.*, PVDF) tribomaterial.

The driver to optimize optoelectronics devices, such as photovoltaic diodes, LEDs and transistors, mainly relies on interface engineering, *i.e.*, the possibility to vary the charge injection or extraction at the electrode and active material interface.[82–84] In TENGs, due to the insulating nature of the tribomaterials and their thickness (few µm), no charge transfer to the electrode can occur, and the triboelectrification charges remain confined in the few nm thick surface layer far from the underlying electrode.[85] A critical aspect in TENGs output performance is therefore the electric field propagation within the tribomaterial due to its dielectric response upon surface triboelectrification.[21] All factors that can vary the displacement field (*D*, eq.1), responsible for the accumulation of the induction charges at the tribomaterial/electrode interface, need to be controlled for reaching the optimal TENG output. To this purpose, also the role played by the energetic at the tribomaterial/electrode interface has to be taken into account. Figure 6-a reports a scheme of the proposed mechanism responsible for the strong improvement that can be obtained upon energetic engineering of the electrode surface in TENGs. When two different work function electrodes ($\phi_1$, $\phi_2$) are placed into electrical contact under thermal equilibrium, a Fermi levels energy alignment occurs, with a consequent flow of charges between the two electrodes.[86] Such phenomenon is well known and is at the base of Kelvin probe measurements.[87-88] In TENGs, this initial situation sets when the two triboelectrodes are placed in contact through the external electronic circuit, before any physical contact is made between the two tribomaterials. Therefore, the presence of two electrodes with different work functions determines the establishment of a built-in potential ($V_{bi}$) and a corresponding electric field ($E_{bi}$) into the TENG. It is intuitive to expect that if such built-in field is opposite in sign to the triboelectrification field generated by the triboelectric charges, an overall decrease of the displacement field should be expected, with a consequent decrease in the TENG output. Vice-versa the built-in field can sum up to the triboelectrification field, giving rise to an overall improvement of the device performance. This simple relation is described in eq. 2, in which the contribution of the electric field (E) to the displacement field D (eq. 1), is expressed in terms of the electric field generated by the triboelectrification charges ($E_{tribo}$) and the preset built-in electric field ($E_{bi}$) established by the electrical contact of the Fermi energy alignment between the two electrodes:

$$D = (E_{tribo} + E_{bi}) + P \qquad (2)$$

Figure 6 represents a schematic view of the aforementioned mechanism. Here, all TENGs are composed by the same tribo-materials (nylon and PVDF membranes can be taken as example), while they vary only for the presence of electrodes with different work functions. Figure 6 -a and -b show the case in which there is no difference in work function between the two electrodes ($\phi_1$ = $\phi_2$). In Figure 6-c and -d the case of a TENG containing two different electrodes is reported, with the lower work function electrode ($\phi_2$) placed at the interface with the more electronegative tribomaterial (tribo2). In Figure 6-e and -f the TENG is constituted by two different electrodes with the lower work function electrode ($\phi_2$) interfacing the more electropositive tribomaterial (tribo1). The Fermi and vacuum levels shifts, shown in Figure 6-a, -c and -e, follow the expected energy level alignment due to the thermodynamic equilibrium, which sets upon electrical contact between the indicated electrodes and which determine a charge flow to the electrode interfaces. Figures 6-c and -e show that a variation of the work function of the two electrodes generates a built-in potential into the TENG, which equals the difference in electrode work function ($V_{bi} \propto \phi_1 - \phi_2$). Figures 6-b, -d and -f, show the built-in potential effect on the charge accumulation in an operating TENG by summing-up (or subtracting) to the triboelectrific field in the air gap ($E_{tribo}$). When the electrode with the lower work function is interfacing the negative tribomaterial, as for the case of the alkylthiolated Au at the PVDF interface, the direction of the field induced by $V_{bi}$ (see also the direction of SAM's dipole moments depicted in Figure 6-d) is equal in sign to the electric field generated upon negative charging due to triboelectrification. Consequently, a stronger electric field is established within the TENG compared to the device having the same electrode work function value (Figure 6-b).

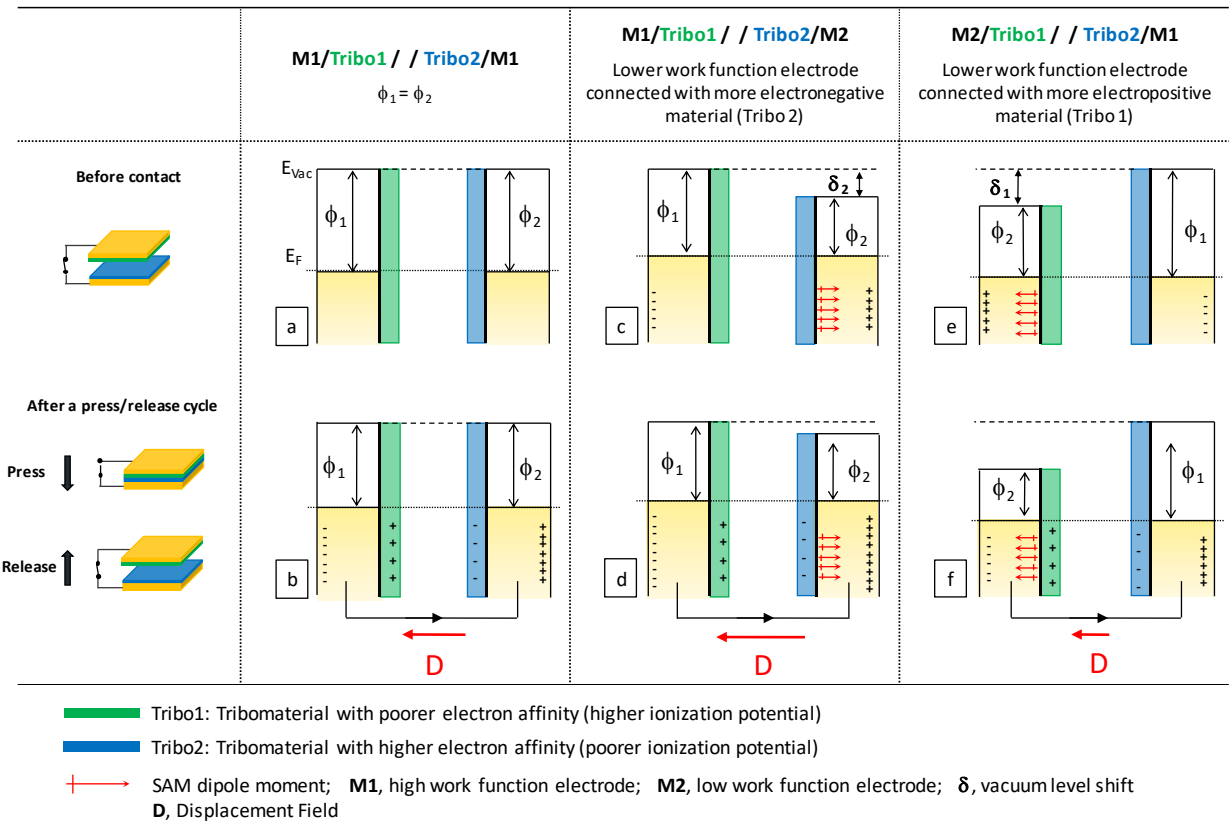

**Figure 6:** Schematic diagram showing the influence of electrodes work function on the TENGs output. The presence of a molecular dipole moment (red arrows) is taken as an example of electrode surface engineering to modify the electrode work function. a), c), e) Fermi levels energy alignment due to short-circuit between the two electrodes in absence of contact electrification. b), d), f) Surface triboelectrification followed by charge flow through the external circuit following a press and release cycle. Red arrows indicate the direction of the dipole moment introduced by an alkylthiolated SAM. In c) the alkylthiol SAM is present only at the interface with the more electronegative tribomaterial, while in e) the SAM is interfacing the more electropositive tribomaterial.

In the configuration reported in figure 6-d) a higher charge density is accumulated at the electrodes compared to the configuration of Figure 6-b), leading to the observed strong improvement in power density output (3.5-fold increase, Figure 4-d). It is therefore intuitive to predict that, if the lower work function electrode is interfacing the positive tribomaterials (as for the case of the alkylthiolate-Au placed at the interface with nylon shown in SI), the field induced by the built-in potential is opposite in sign to the electric field generated upon triboelectrification. This condition determines a decrease in the accumulated charge density.

A simple electrodynamic model of the working potential in a contact-separation TENG, can be drawn under the assumption of infinitely large electrodes as previously reported by Niu et al.[89]

The different contributions to the working potential (V) of the TENG are described in the scheme reported in Figure 7 and SI-7. $E_{tribo}$ is the electric field present within the air gap generated upon triboelectrification of the two tribomaterials, tribo1 and tribo2, owing a charge density $+\sigma$ and $-\sigma$, respectively.

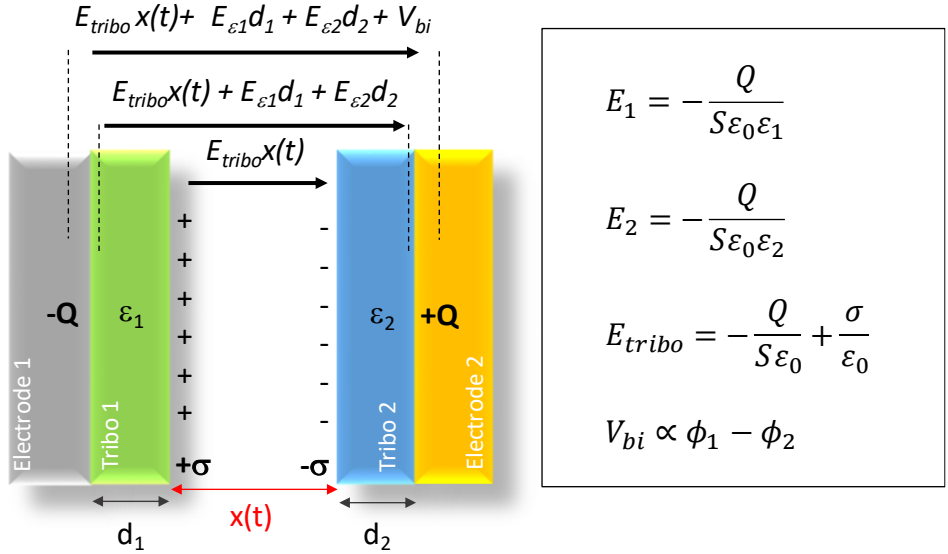

**Figure 7**: Scheme of the different contribution to the working potential (V, eq. 3) in a TENG operating under contact separation mode. $E_{tribo}$, triboelectrification field in the air gap; $E_1$ and $E_2$ electric field in the dielectric 1 and 2 respectively; $V_{bi}$ built-in potential.

$E_{\varepsilon 1}$ and $E_{\varepsilon 2}$ are the electric field within the dielectric layers (tribo1, dielectric contant $\varepsilon_1$; tribo2 with dielectric constant $\varepsilon_2$) due to the accumulation of charges at the electrodes 1 (-Q) and 2 (+Q).

As described above an additional contribution must be considered for the presence of a built-in potential ($V_{bi}$) due to the difference in electrodes work functions ($\phi_1$ and $\phi_2$ are the work function associated to electrode 1 and electrode 2, respectively). Given the distances $d_1$ (thickness of tribo1), $d_2$ (thickness of tribo2) and x(t) (relative distance between tribo1 and tribo2 surfaces varying with time) the TENG working potential (V) can be described as follow:

$$V = E_1 d_1 + E_2 d_2 + E_{tribo} x(t) + V_{bi} \qquad (3)$$

From eq. 3 we can derive the open circuit potential ($V_{oc}$) and the charge generated under short circuit conditions ($Q_{sc}$), which are expressed as follow (see SI):

$$V_{oc} = \frac{\sigma}{\varepsilon_0} x(t) + V_{bi} \qquad (4)$$

$$Q_{sc} = (\sigma x(t) + V_{bi}\varepsilon_0) \frac{S}{d_0 + x(t)} \qquad (5)$$

in which $d_0$ is the effective distance defined as $d_0 = \frac{d_1}{\varepsilon_1} + \frac{d_2}{\varepsilon_2}$.

From this simple model, it is evident that by taking into account the direction of the built-in field given by the difference in electrodes work function, the contribution of the $V_{bi}$ to the $V_{oc}$ and $I_{sc}$ can lead to either a decrease or an increase in the measured output values.

It is now clear that the design of a new TENG architecture has to consider the most appropriate match between electrode work function and the triboelectric material. In particular: 1) a wide difference in the electrodes work function leading to high $E_{bi}$ can increase the power output; 2) in order to benefit of the positive effect of a favorable $E_{bi}$, the lower work function electrode should interface with the more electropositive tribomaterial, while the more electronegative tribomaterial has to interface the higher work function electrode. Since the difference in work function between the Au-CH$_3$ electrode and the as-produced FLG electrode is only 0.17 eV, we also highlight the important role played by the contact area between the electrode surface and the tribomaterial. In particular, increasing this interfacial area can boost the power efficiency due to an increase of the overall capacitance of the TENGs.

### 4. Conclusion

In this work, we prove that FLG electrodes represent a highly efficient charge collector in TENGs, overcoming the performances of conventional metal electrodes, with the added value of easy integration into flexible devices via solution processing in ambient conditions. We demonstrate that the design of new TENG configurations requires the complete understanding of all factors that affect the triboelectric charging and induction processes, which have to take into account not only the charge affinity and triboelectric properties of the tribomaterials but also the processes occurring at the tribomaterial/electrode interfaces. Here, we demonstrate that the work function of the electrodes is a fundamental parameter to be controlled for the generation of a favorable built-in potential in TENGs. This built-in potential contribute to enhance the electrical performances by increasing the charge density accumulated at the electrode/tribomaterial interface. In addition, our finding also bring to the conclusion that the selection of the electrode can modify the relative positioning of the tribomaterial along the triboelectric series. Therefore, this parameter has to be taken into account when designing experiments aiming at defining the tendency of insulating materials to undergo triboelectrification.

 Acknowledgements

The authors thanks the Electronic Design Laboratory and the Mechanical Workshop teams at IIT for discussion and technical support. This project has received funding from the European Union's Horizon 2020 research and innovation program under grant agreement no.785219-GrapheneCore2.

# Electrode selection rules for enhancing the performance of triboelectric nanogenerators and the role of few-layers graphene

G. Pace[1], A. Ansaldo[1,2], Michele Serri,[1] Simone Lauciello,[1] and F. Bonaccorso[1,3]

[1]*Istituto Italiano di Tecnologia, Graphene Labs, Via Morego, 30, 16136 Genova, Italy*
[2] *ASG Superconductors S.p.A., corso Ferdinando Maria Perrone, 73R, 16152 Genova, Italy*
[3] *BeDimensional S.p.A., via Lungotorrente Secca 3d, 16163 Genova, Italy*

Corresponding authors: *giuseppina.pace@iit.it*,
*francesco.bonaccorso@iit.it*

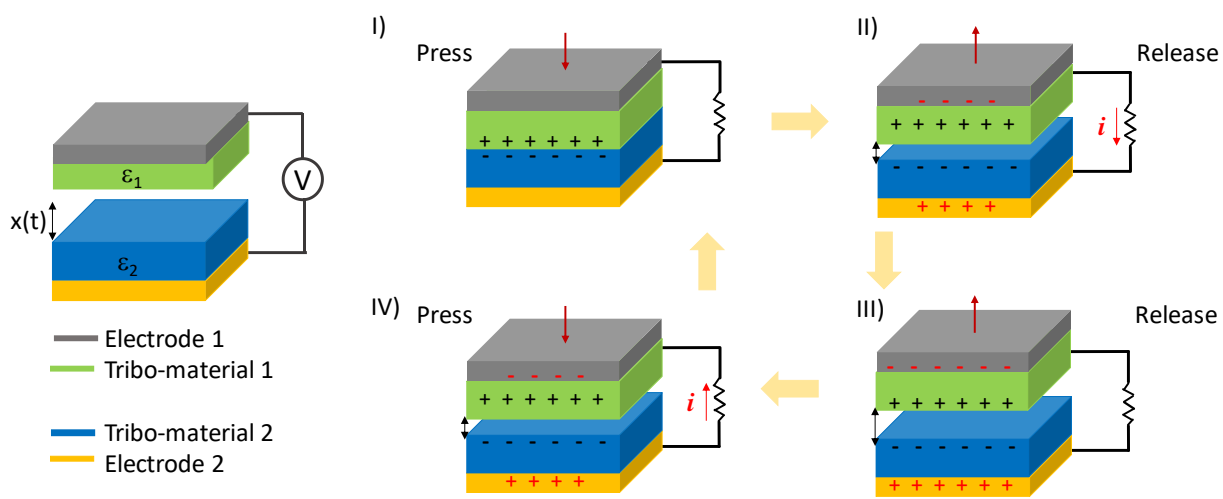

**Figure SI-1:** Scheme of a TENG working in contact-separation mode.

**Figure SI-1** shows the working mechanism of a triboelectric nanogenerator (TENG) operating in vertical contact-separation mode. At each press and release event, a current flow through the external circuit. $\varepsilon_1$ and $\varepsilon_2$ are the dielectric constants of the tribomaterial 1 and the tribomaterial 2, respectively, and x(t) is the time dependent separation distance between the two triboelectrodes.

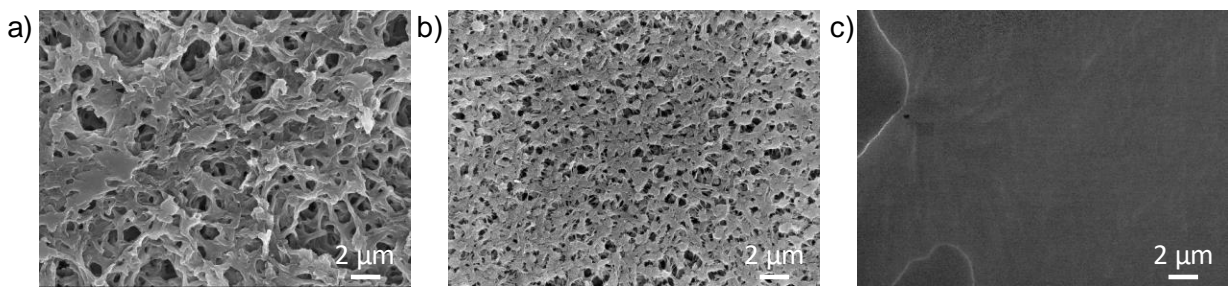

**Figure SI-2**: SEM images of the insulating polymer membranes used as triboelectric materials in the Tengs: a- Nylon; b- PVDF; c- polyurethane.

Figure SI-2 presents the scanning electron microscopy (SEM) images of the top of the commercial porous polyvinylfluoride (PVDF) and Nylon membranes that have been employed in this work. Polyurethane was employed to acquire the cross section images of the few-layer graphene (FLG)-based flexible electrodes presented in Figure 2 of the main text.

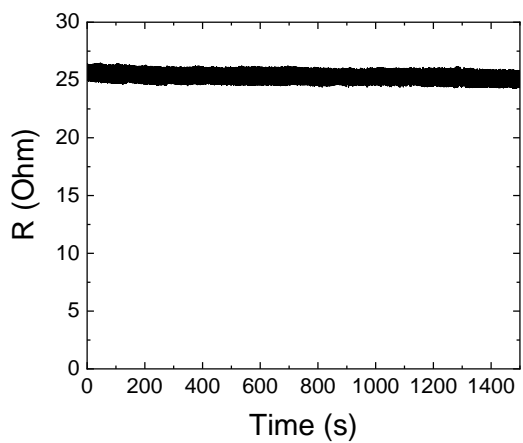

**Figure SI-3:** Resistance measured across the FLG flexible electrodes under continuous bending cycles.

Measurements in Figure SI-3 were acquired on a 4 × 4 cm$^2$ FLG flexible electrodes as the one used for fabricating the TENGs devices in our work. During the bending, the curvature radius was varied from 2 cm to 1 cm at a frequency of ~ 2Hz (**Movie 1**). The memory buffer of our measurement set-up dictates the time scale in the figure. The electrode can be cycled over at least two days with no variation of the electrical properties.

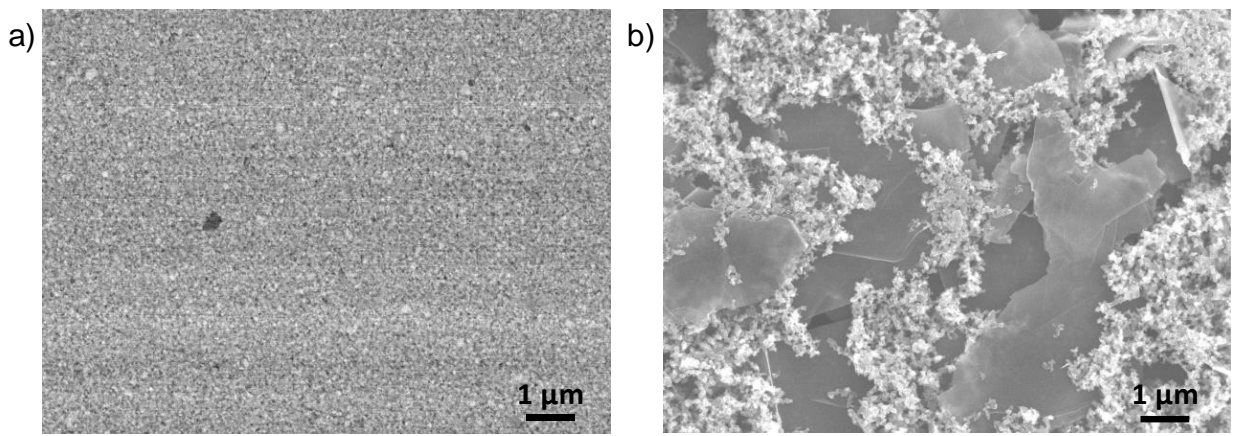

**Figure SI-4:** SEM images of the electrode surfaces: a- gold electrode; b- FLG flexible electrode.

Figure SI-4 shows the top view of the electrodes used in this work for the fabrication of the triboelectrodes. Figure SI-4 shows the presence of the FLG and of the C-nanoparticles agglomerates that protrudes from the electrode surface.

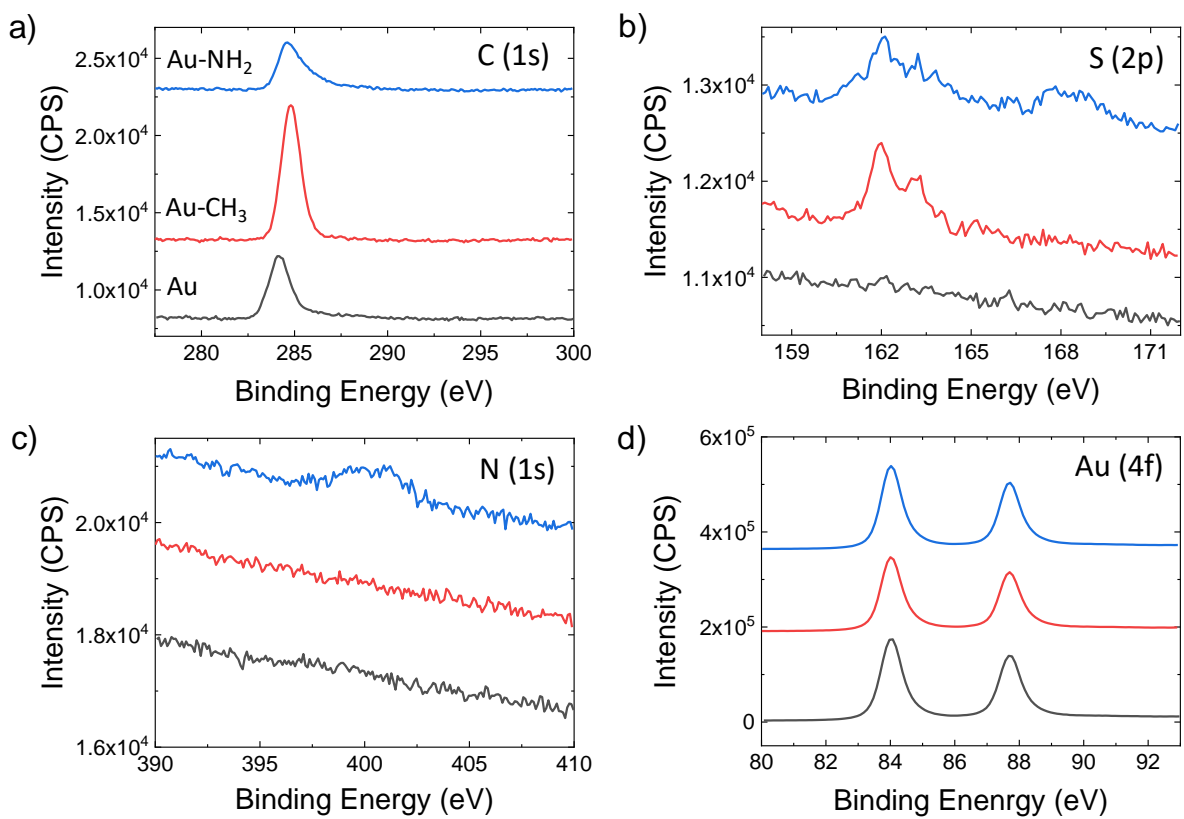

**Figure SI-5:** XPS narrow spectra of bare Au (grey) and Au functionalized with 1-dodecanethiol SAM (Au-CH$_3$) and 1-aminoethanethiol SAM (Au-NH$_2$). A- C (1s) spectra; B- S (2p) spectra; C- N (1s) spectra; D- Au (4f) spectra.

The X-ray photoelectron spectroscopy (XPS) data reported in Figure SI-5 show the effective functionalization of the Au electrodes with the self-assembled monolayers (SAMs). The presence of the S(2p) peak demonstrates the presence of the thiolated SAM on both the 1-dodecanethiol SAM (Au-CH$_3$) and 1-aminoethanethiol SAM (Au-NH$_2$). A pronounced C(1s) peak is observed for the dodecanthiol SAMs, in which 12C a time are present per each alkylic chain. A clear N(1s) is present for the 1-aminoethanthiolated SAM as expected.

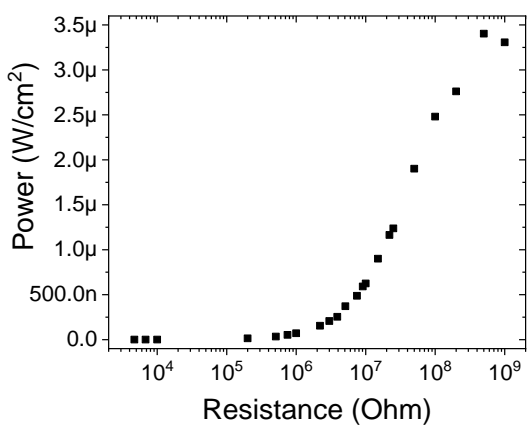

**Figure SI-6:** Power output of a TENG embedding aluminum electrodes (Al/Nylon // PVDF/Al).

Aluminum electrodes are reported to have a work-function in the range of 4.06 - 4.41 eV.[1] Such a value is closer to the work-function found for Au-CH$_3$ functionalized gold function (4.37 eV). Consistently, the power maximum found for TENGs embedding Al electrodes (3.40 µW/cm$^2$) is smaller than what found for the TENG embedding the Au-CH$_3$ electrode (3.78 µW/cm$^2$), but higher than the power density found for the Au-NH$_2$ electrode (work function 4.66 eV and 1.57 µW/cm$^2$). These data further confirm the TENGs power density dependence on the electrodes work-function.

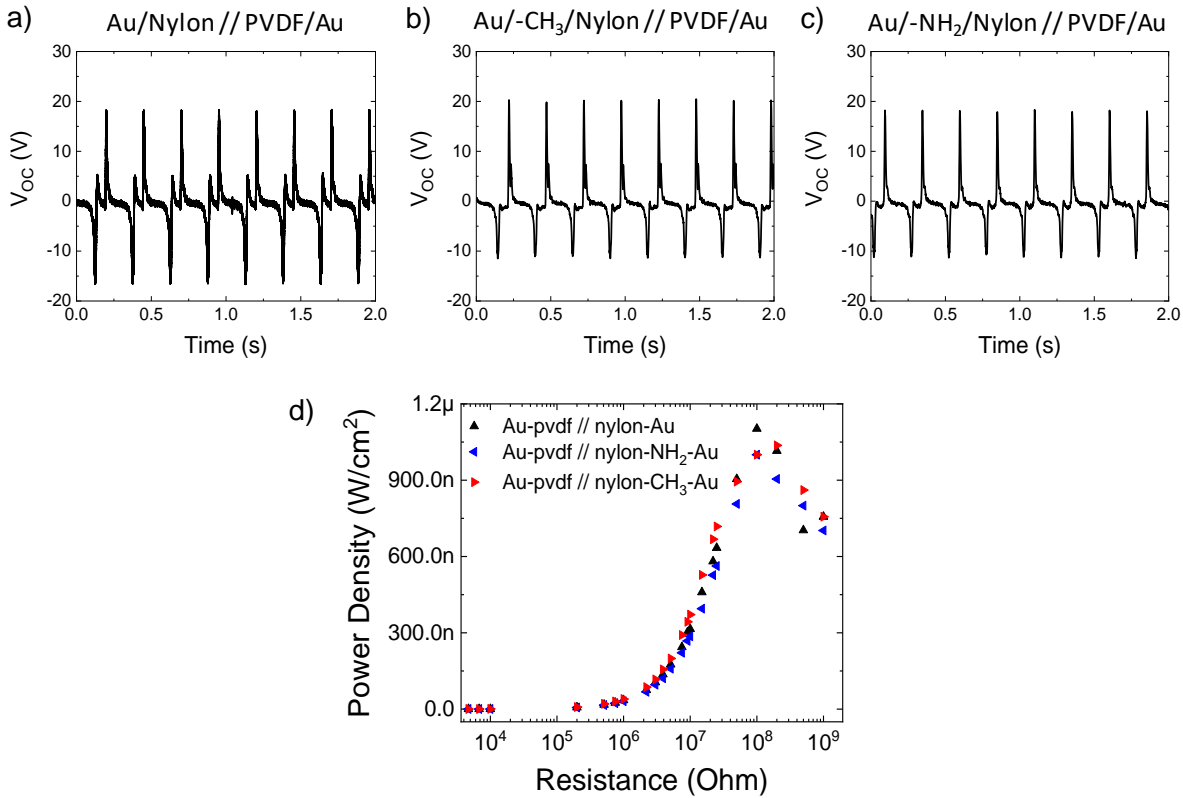

**Figure SI-7:** Open circuit voltage ($V_{OC}$) outputs and power density generated by TENGs fabricated with gold electrodes functionalized with different SAMs: a) no SAM on Au; b) 1-dodecanethiol SAM on Au; c) 1-aminodecanthiol SAM on Au. (40MΩ probe, 4Hz, 10N, 5mm)

Figure SI-7 shows the data acquired on TENGs, in which different functionalized gold electrodes are interfaced with the nylon membrane, while a bare gold electrode is present on the PVDF side. Differently from the case presented in Figure 5 of the main text, we here observe a decrease of power density when the SAMs are present on the Au surface.

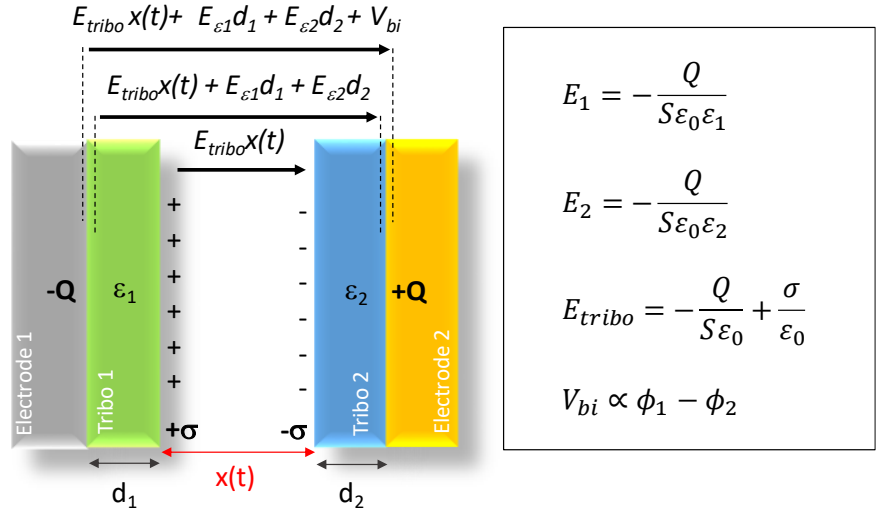

**Figure SI-8:** Sketch of the different component to the TENG potential difference between two triboelectrodes in a contact-separation configuration.

The area of the electrodes (S) is much larger than the distance between the two triboelectric surfaces and a simple model based on the assumption of infinitely large electrodes can be drawn. Similarly, to the analytical model presented by Niu *et al.*,[2] we can define the following electric fields according to basic electrodynamics principles:

$$E_1 = -\frac{Q}{S\varepsilon_0\varepsilon_1} \tag{1}$$

$$E_2 = -\frac{Q}{S\varepsilon_0\varepsilon_2} \tag{2}$$

$$E_{Air} = -\frac{Q}{S\varepsilon_0} + \frac{\sigma}{\varepsilon_0} \tag{3}$$

Equation 1, 2, 3 define the electric field within the tribomaterial 1 ($E_1$, dielectric material 1, with dielectric constant $\varepsilon_1$), within the tribomaterial 2 ($E_2$, dielectric material 2, with dielectric constant $\varepsilon_2$) and the electric field within the air gap between the two tribomaterials 1 and 2. Q is the amount of charges exchanged between the two electrodes, while $\sigma$ is the charge density present at the two tribomaterials surface following contact electrification. S is also the area of the triboelectric surfaces, and *x(t)* is the variable distance between the two tribomaterials.

In addition to the electric potential established by the field described above, our work shows that in order to predict the effect of the selection of the electrodes on the TENGs performance, an additional contribution from the field generated by the electrode work function dependent built-in potential ($V_{bi}$) has to be considered. Such $V_{bi}$ is a function of the difference in work-function of the two electrodes ($\phi_1$ and $\phi_2$).

$$V_{bi} \propto \phi_1 - \phi_2 \qquad (4)$$

Introducing the $V_{bi}$, we can describe the voltage between the two electrodes as:

$$V = E_1 d_1 + E_2 d_2 + E_{tribo} x(t) + V_{bi} \qquad (5)$$

Taking into account eq. 1-4, we find:

$$V = -\frac{Q}{S\varepsilon_0}\left(\frac{d_1}{\varepsilon_1} + \frac{d_2}{\varepsilon_2} + x(t)\right) + \frac{\sigma}{\varepsilon_0}x(t) + V_{bi} = -\frac{Q}{S\varepsilon_0}(d_0 + x(t)) + \frac{\sigma}{\varepsilon_0}x(t) + V_{bi} \qquad (6)$$

In which $d_0 = \frac{d_1}{\varepsilon_1} + \frac{d_2}{\varepsilon_2}$ is the effective thickness constant.

Under open circuit conditions, in which no charges are allowed to flow in the external circuit (Q=0) the open circuit voltage ($V_{oc}$) is expressed as follow:

$$V_{oc} = \frac{\sigma}{\varepsilon_0}x(t) + V_{bi} \qquad (7)$$

Therefore, differently from the previously reported literature,[2] the $V_{oc}$ is not only dependent on the distance dependent potential established between the two tribomaterials as a consequence of the contact electrification ($V_{tribo} = \frac{\sigma}{\varepsilon_0}x(t)$) but also on the $V_{bi}$.

The charges exchanged under short circuit conditions ($Q_{sc}$) can be derived imposing V=0 and as follow:

$$0 = -\frac{Q}{S\varepsilon_0}(d_0 + x(t)) + \frac{\sigma}{\varepsilon_0}x(t) + V_{bi} \qquad (8)$$

and

$$Q_{sc} = (\sigma x(t) + V_{bi}\varepsilon_0)\frac{S}{d_0 + x(t)} \qquad (9)$$

From equation 9 we can derive the short circuit current:

$$I_{sc} = \frac{dQ}{dt} = \frac{S}{d_0+x(t)}\left[\sigma - \frac{\sigma x(t)+V_{bi}\varepsilon_0}{d_0+x(t)}\right]\frac{dx(t)}{dt} \qquad (10)$$

Equation 7 and equation 10 show the contribution of the $V_{bi}$ to the two most important figures of merit in TENGs ($V_{oc}$ and $I_{sc}$).

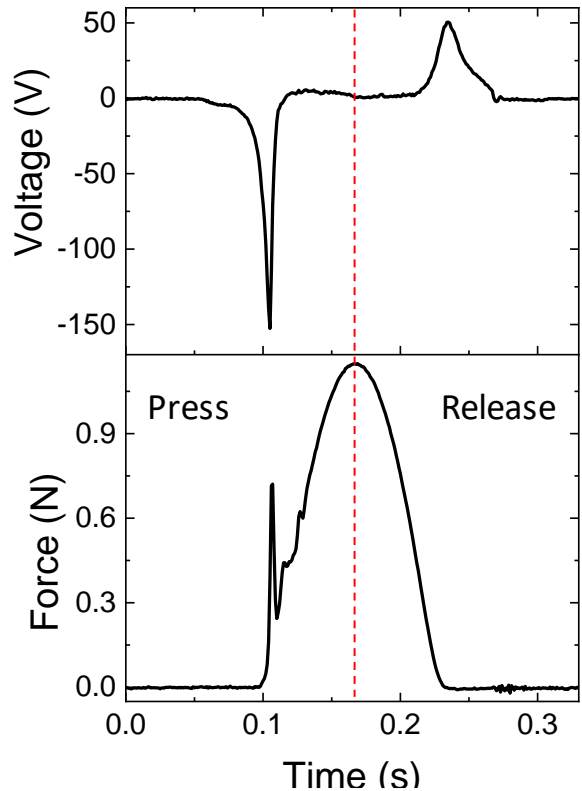

**Figure SI-9**: Voltage and force curve recorded during a press and release cycle for a TENG operating in contact separation mode (40MΩ probe, 3Hz, 10N, 3mm).

Figure SI-9 shows the force and voltage time profile of a TENG, which contains FLG flexible electrodes both as bottom and top electrodes. Polyvinylfluoride and Nylon membranes are used as negative and positive tribomaterials, respectively. A voltage peak is observed at each press and release event.